\documentclass[]{iopart}  

\usepackage{epsfig}
\newcommand{\be}[1]{\showlabel{#1}\begin{equation}\label{#1}}
\newcommand{\ee}{\end{equation}}   
\newcommand{\bea}{\begin{eqnarray}}
\newcommand{\eea}{\end{eqnarray}} 
\newcommand{\eq}[1]{(\ref{#1})}
\newcommand{\fig}[1]{figure~\ref{#1}} 

\newcommand{\ba}[1]{\begin{array}{#1}}
\newcommand{\ea}{\end{array}}   

\newcommand{\maxi}{\mbox{\scriptsize max}}	
\newcommand{\mini}{\mbox{\scriptsize min}}	

\newcommand{\showlabel}[1]{}

\usepackage[colorlinks,hypertex]{hyperref}

\begin{document}

\paper[Synchronization in the "Zajfman trap" II]
{A mapping approach to synchronization in the "Zajfman trap". II: 
the observed bunch}

\author{Tiham\'{e}r\ Geyer\dag\ 
		and David\ J\ Tannor\ddag}
\address{\dag\ Zentrum f\"ur Bioinformatik, 
		Universit\"at des Saarlandes, D--66041 Saarbr\"ucken, Germany}
\address{\ddag\ Department of Chemical Physics, Weizmann Institute of
		Science, Rehovot 76100, Israel} 

\begin{abstract}
	We extend a recently introduced mapping model, which explains the
	bunching phenomenon in an ion beam resonator for two ions [Geyer,
	Tannor, \emph{J. Phys. B} \textbf{37} (2004) 73], to describe the
	dynamics of the whole ion bunch. We calculate the time delay of
	the ions from a model of the bunch geometry and find that the
	bunch takes on a spherical form at the turning points in the
	electrostatic mirrors. From this condition we derive how the
	observed bunch length depends on the experimental parameters. We
	give an interpretation of the criteria for the existence of the
	bunch, which were derived from the experimental observations by
	Pedersen \etal [Pedersen \etal, \emph{Phys.~Rev.~A} \textbf{65}
	042704].
\end{abstract}


\submitto{\jpb}

\vspace{5mm}

\section{Introduction}

Ions in an ion trap usually behave as a gas: they try to fill all the
available volume, defined by the trap's electromagnetic fields. It was
only in storage rings where the ions could be forced into well
localized bunches. But even there the ions' mutual repulsion and the
inevitable spread in their kinetic energies tend to distribute the
ions throughout the whole ring.

Recently a surprising cooperative behavior of the ions was discovered
in a Zajfman trap \cite{PED01}, an ion trap resonator built of only
two opposite electrostatic mirrors and two focussing lenses
\cite{ZAJ97}: the ions are injected into the trap from an ion beam and
this injected bunch of ions bounces back and forth between the two
mirrors but does not diffuse in the trap. The ions, which all have the
same charge and also slightly different energies and trajectories
through the trap, seem to ``stick together''; they are not only
confined to the trap volume, but also seem to be trapped within the
bunch, which oscillates back and forth like a single macroscopic
``super particle''. This bunch works as a ``dynamic inner trap'',
itself moving between the static electric fields of the external trap.

As a first application of this ``Zajfman trap'' operating in the
bunching regime a high resolution Fourier transform mass spectrometer
has been demonstrated with a resolution of $\Delta m/m \approx 7
\times 10^{-6}$ \cite{STR02}. This value was until now only achieved
in storage rings \cite{MAR98}. This high resolution is made possible
by observing the oscillations of the macroscopic bunch through the
trap over time intervals of several tenths of a second. The
observation time is limited by the trapping time of the ions in the
inner trap (the bunch), which is determined by the number of
collisions between the ions and residual gas atoms in the vacuum
chamber \cite{ATT05x}.

The operating regime of the trap and the stability conditions, under
which this counter intuitive ``self--bunching'', or synchronization,
occurs, were first explained by Strasser \etal \cite{STR02} by relating
the ion dynamics to the so called negative mass instability
\cite{LAW88}: due to the special dispersion of the trap the period of
an ion increases with its energy.  This can be described by a negative
effective mass of the ions.  Consequently the ions have to repel each
other in order to synchronize their motion.

A microscopic explanation of the observed synchronization for two
identical ions was given by the authors in reference \cite{GEY03}
which we refer to as ``part I'' in the following. There the problem is
described by a set of simple mappings for the evolution of the ions'
relative coordinate through the different parts of the trap.  The ions'
interaction is incorporated phenomenologically as a time delay.  With
this approach we confirm not only the already known stability
conditions but describe the microscopic process taking place:
synchronization is a continued alternation of energy and position
exchange between the two ions.

In this paper we show how the model of two identical ions can be
generalized to describe the dynamics of the whole bunch. For this we
associate one of the ions with a ``test ion'', which moves through the
cloud of the other ions, and the other describes the motion of the
bunch's CM. With the two identical ions we did not need to specify the
interaction to explain the mechanism, but now the interaction
potential that the test ion feels is an important part of the effect:
it is derived from the charge distribution in the bunch, i.e., the
test ion's average position in the bunch, which in turn is determined
selfconsistently from the motion of the test ion inside this bunch.
This test ion case is a specialization as well as a generalization of
the two identical ions description: by specifying the interaction we
can describe the generalized behavior of arbitrarily many ions through
a mean field treatment.

The paper is organized as follows: Section \ref{sec:twoIons} reviews
the mapping approach of part I. We show that the approach can be
extended from the two--ion case to $N$ ions. The key issue left
unresolved in Section \ref{sec:twoIons} is how to calculate the time
delay of the test ion in the bunch. This question is taken up in
Section \ref{sec:EstimateLength}, where we show how this parameter can
be obtained from a mean field treatment of the bunch potential. A
central result of Section \ref{sec:EstimateLength} is the derivation of
the spherical geometry of the bunch at the turning points. In section
\ref{sec:DelayBehave} the dependence of the bunch length on the
various parameters of the trap and the ions is derived. In section
\ref{sec:PhaseSpace} the momentum spread of the bunch is
calculated. Then, in section \ref{sec:Criteria} we show how the
experimental criteria for bunching deduced from the observations by
Pedersen \etal \cite{PED02b}, are connected to our explanations. The
results are summarized and further developments are sketched in
section \ref{sec:summary}.

\section{Review of the mapping model}
\label{sec:twoIons} \showlabel{sec:twoIons}

\subsection{Coordinate system, transit times and the dispersion relation}

In this section we briefly review the basics of the two ion case. For
information about the trap and the experimental findings the reader is
referred to references \cite{PED02b} and \cite{PED02a}. For more
details about the mapping model we refer the reader to ``part I''
\cite{GEY03}.

Consider two ions with the simplified trap potential:
\be{eq:trapPotential}
	V(x) = \left\{ 
	\begin{array}{ccl}
		0                      & \mbox{ when } & |x| \leq \frac{L}{2}  
			\\[0.2cm]
		(|x| - \frac{L}{2})\, F & \mbox{ when } & |x| > \frac{L}{2}\, .
	\end{array}
	\right.
\ee
$F$ is the constant gradient of the mirror fields, $L$ is the field
free distance between the two mirrors.

In general the two ions have masses $m_1$ and $m_2$ and charges $q_1$
and $q_2$, respectively. As synchronization shows up in the distance
between the ions, we use center of mass (CM) and relative coordinates.
With the total mass $M = m_1 + m_2$ and the reduced mass $\mu =
\frac{m_1 m_2}{m_1+m_2}$ we define
\be{eq:CmRelCoordsAllg}
	\left.
	\begin{array}{c}
		R = \frac{m_1 x_1 + m_2 x_2}{M}  \\[0.1cm]
		x = x_1 - x_2
	\end{array}
	\right\} \Leftrightarrow \left\{
	\begin{array}{c}
		x_1 = R + \frac{m_2}{M} x \\[0.1cm]
		x_2 = R - \frac{m_1}{M} x \, .
	\end{array}
	\right.
\ee
The corresponding momenta are
\be{eq:CmRelMomentaAllg}
	\left.
	\begin{array}{c}
		P = p_1 + p_2  \\[0.1cm]
		p = \frac{m_2}{M}p_1 - \frac{m_1}{M}p_2
	\end{array}
	\right\} \Leftrightarrow \left\{
	\begin{array}{c}
		p_1 = \frac{m_1}{M} P + p \\[0.1cm]
		p_2 = \frac{m_2}{M} P - p \, .
	\end{array}
	\right. 
\ee
$R$ and $P$ are the CM coordinate and momentum, respectively; the
relative motion is described by $x$ and $p$.  The ions' positions and
momenta are denoted by $x_1$, $p_1$ and $x_2$, $p_2$.

With the ion--ion interaction $W(x)$ the Hamiltonian for these two
ions reads
\showlabel{eq:Hamiltonian}
\bea
	H & = & \frac{p_1^2}{2m_1} + \frac{p_2^2}{2m_2} + 
		q_1 V(x_1) + q_2 V(x_2) + q_1 q_2 W(x_1-x_2) \nonumber \\
	& = & \frac{P^2}{2M} + \frac{p^2}{2\mu} + 
		q_1 V(R+\textstyle{\frac{m_2}{M}}x) + 
		q_2 V(R-\textstyle{\frac{m_1}{M}}x) + q_1 q_2 W(x) \, .
	\label{eq:Hamiltonian}
\eea
To separate CM and relative coordinates when both ions are in the same
part of the trap potential we need that $q_1V(x_1) + q_2 V(x_2) =
(q_1+q_2)V(R)$. With the potential of equation \eq{eq:trapPotential}
we get the condition that both ions must have the same charge to mass
ratio:
\be{eq:ChargeMass}
	\frac{q_1m_2 - q_2m_1}{M} = 0 \quad \Leftrightarrow \quad
	\frac{q_1}{m_1}=\frac{q_2}{m_2}
\ee
For two identical ions this condition is trivially fulfilled. Here we
describe a bunch of $N+1$ identical ions; then the above condition is
also fulfilled, if we take one of the ions to be a test ion of mass
$m$ and charge $q$ and identify the other with the remaining $N \geq
1$ identical ions of the bunch:
\bea
	m_1 = m & \quad & m_2 = Nm \\
	q_1 = q & \quad & q_2 = Nq
\eea
The number of ions $N$ is consequently a positive integer, which can
be $N=1$, too. Now, with $M = (N+1)m$ and $\mu = \frac{N}{N+1}m$, the
CM and relative coordinates (equations \eq{eq:CmRelCoordsAllg} and
\eq{eq:CmRelMomentaAllg}) become:
\be{eq:CmRelCoords}
	\left.
	\begin{array}{c}
		R = \frac{x_1 + N x_2}{N+1}  \\[0.1cm]
		x = x_1 - x_2
	\end{array}
	\right\} \Leftrightarrow \left\{
	\begin{array}{c}
		x_1 = R + \frac{N}{N+1} x \\[0.1cm]
		x_2 = R - \frac{1}{N+1} x
	\end{array}
	\right.
\ee
and
\be{eq:CmRelMomenta}
	\left.
	\begin{array}{c}
		P = p_1 + p_2  \\[0.1cm]
		p = \frac{N}{N+1}p_1 - \frac{1}{N+1}p_2
	\end{array}
	\right\} \Leftrightarrow \left\{
	\begin{array}{c}
		p_1 = \frac{1}{N+1} P + p \\[0.1cm]
		p_2 = \frac{N}{N+1} P - p
	\end{array}
	\right. .
\ee

In part I \cite{GEY03} we combined the geometry of the trap and the
ions' mass and energy into two parameters, which are derived from the
times that the ions spend in the mirror, i.e., at $|x| > L/2$, and in
the central part of the trap, respectively.

An ion with the laboratory energy $E_0$ and the momentum
$p_0=\sqrt{2mE_0}$ spends the time $T_m$ inside the mirror:
\be{eq:TimeInTheMirror}
	T_m = \frac{2p_0}{qF} = \frac{2P}{NqF}
\ee
$T_m$ is one of the two parameters mentioned above. 

When the ion turns around at $T_m/2$ it has penetrated into the mirror
potential for the distance
\be{eq:XmDef}
	X_m = \frac{E_0}{qF}.
\ee
The ions have the velocity $p_0/m \approx P/M$ and therefore need the
time
\be{eq:TimeInTheFlat}
	T_f = \frac{Lm}{p_0} = \frac{LM}{P} := \alpha T_m
\ee
to pass through the central field free region of the trap.

It turns out that not the times themselves but their ratio $\alpha$ ---
the second parameter --- determines, if synchronization occurs.  When
we express $\alpha$ in terms of the experimental parameters, where $U$
is the acceleration voltage, i.e. $E_0 = qU$, we get:
\be{eq:AlphaExp}
	\alpha = \frac{L\,qF}{4E_0} = \frac{LF}{4U},
\ee
With $T_m$ and $T_f$ we calculate the total time $T$ for one period as
\showlabel{eq:TimeForPeriod}
\bea
	T & = & 2T_m + 2T_f =
		\label{eq:TimeForPeriod} \\
	& = & \sqrt{\frac{2m}{q}}
		\left( \frac{4\sqrt{U}}{F} + \frac{L}{\sqrt{U}} \right)
\eea
This form highlights the importance of the bunching effect for mass
spectrometry applications: the period of the bunch, i.e., the
synchronized ions, which can be measured very precisely for a huge
number of revolutions, is directly related to the ratio $\frac{m}{q}$.

The factor $\alpha$ also shows up in the dispersion $\frac{\partial
T}{\partial p_0}$ of the trap, i.e., in the dependence of the period of
one ion on its momentum.  With equations \eq{eq:XmDef} and
\eq{eq:AlphaExp} the dispersion is a simple relation between the
``geometric'' parameters:
\be{eq:Dispersion}
	\frac{\partial T}{\partial p_0} = \frac{4}{F}(1-\alpha) = 
		\frac{1}{U}\left(4X_m-L\right)
\ee
For $(1-\alpha) > 0$ the period $T$ increases with the ion's momentum 
$p_0 = \sqrt{2mE_0}$ and vice versa.

\subsection{Independent ions: The momentum kick and the basic mapping structure}

In our model potential the CM and the relative motion separate for
nearly all the time, except for a short interval $\tau_k$. This is the
time between when the first ion crosses the kink between the flat part
and one of the mirrors at $\pm \frac{L}{2}$ and when the second one
passes that point, too. During that time the ions are accelerated
relative to each other by the trap potential. For the case of ions
going into the mirror at $L/2$ it is:
\be{eq:tauKick}
	\tau_k = \frac{x}{p_2 / m}
\ee
After this time the momentum $p_2$ of the second ion will be the same,
$p_2'=p_2$, but $p_1$ is slowed down to
\be{eq:p1AfterTau}
	p_1' = p_1 + \tau_k \left( -\frac{\partial V}{\partial x_1} \right) 
         = p_1 - \frac{x m_2}{p_2} qF \, .
\ee
The two ions have almost the same velocity, so $\frac{p_2}{m_2}$ is
approximated by $\frac{p_0}{m}$. The relative momentum $p =
\frac{N}{N+1}p_1 - \frac{1}{N+1}p_2$ \eq{eq:CmRelMomenta} now changes
with equation \eq{eq:TimeInTheMirror} as
\be{eq:MomentumKick}
	p' = p - \frac{Nm}{N+1} \frac{qF}{p_0}x = p - \frac{2\mu}{T_m}x
		\, .
\ee
As $\tau_k$ is short, the relative distance $x$ does not change during
this time in our approximation; also the ions' very weak interaction
can be neglected \cite{GEY03}.  This sudden change of the momentum
\eq{eq:MomentumKick} can be formulated as a mapping $\mathcal{K}$ of
the relative distance and momentum $x$ and $p$ just before this kick
onto the values $x'$ and $p'$ right after the kick:
\be{eq:MappKick}
	\mathcal{K}: 
	\left( \begin{array}{c} x \\ p \end{array} \right) \mapsto
	\left( \begin{array}{c} x' \\ p' \end{array} \right) = 
	\left( \begin{array}{c} x  \\ p - \frac{2 \mu}{T_m}x \end{array} \right)
	.
\ee

Without the ions' interaction the relative coordinate evolves freely
in the central part and inside the mirrors, leading to the two
mappings $\mathcal{F}$ and $\mathcal{M}$, respectively, which differ
only in the length of their time interval:
\be{eq:MappFree}
	\mathcal{F}: 
	\left( \begin{array}{c} x \\ p \end{array} \right) \mapsto
	\left( \begin{array}{c} x' \\ p' \end{array} \right) = 
	\left( \begin{array}{c} x + \frac{\alpha T_m}{\mu}p \\ p \end{array} \right)
\ee
\be{eq:MappMirror}
	\mathcal{M}: 
	\left( \begin{array}{c} x \\ p \end{array} \right) \mapsto
	\left( \begin{array}{c} x' \\ p' \end{array} \right) = 
	\left( \begin{array}{c} x + \frac{T_m}{\mu}p \\ p \end{array} \right)
\ee

The complete dynamics of two noninteracting ions in their CM system
can be described by these three mappings $\mathcal{K}$, $\mathcal{F}$
and $\mathcal{M}$. They consequently form the foundation of our
mapping model. With the ions' interaction incorporated into this
picture we are able to determine the stability conditions for bunching
and, as we will show in this paper, also describe the extension of the
whole bunch with respect to the external parameters.

\subsection{Including the ion--ion interaction: the time delay and the stability condition}

As shown in part I \cite{GEY03}, the ions' motion is synchronized by
their repulsion when the ions interact in the mirror. Without further
specifying the form of the interaction potential we can model its
effect by a time delay $\tau_m$ that affects the relative motion when
the ions' paths cross each other. We include this delay $\tau_m$ into the
mirror mapping as
\be{eq:MappMirrorTau}
	\mathcal{M}': 
	\left( \begin{array}{c} x \\ p \end{array} \right) \mapsto
	\left( \begin{array}{c} x' \\ p' \end{array} \right) = 
	\left( \begin{array}{c} x + \frac{T_m-\tau_m}{\mu}p \\ p \end{array} \right)
	\, .
\ee
Now we can determine the stability of the motion by calculating the
eigenvalues of the linearized mapping
\be{eq:P12}
	\mathcal{P}'_{1/2} = \mathcal{F}_{1/2} \otimes \mathcal{K} \otimes 
		\mathcal{M}' \otimes \mathcal{K} \otimes \mathcal{F}_{1/2}
\ee
of a half period.  They have the form
\be{eq:LambdaP12}
	\lambda_{1/2} = -1 + 2\gamma \pm 2 \sqrt{\gamma^2 - \gamma}\, ,
\ee
where we applied the abbreviation
\be{eq:GammaDef}
	\gamma = \epsilon(1-\alpha) \quad \mbox{with} \quad 
	\epsilon = \frac{\tau_m}{T_m}
\ee
Equation \eq{eq:LambdaP12} was one of the central equations of part I. 
Analyzing the two ion case in this paper it will take on a new
significance in describing the stability of the N--ion bunch.

The three mappings $\mathcal{K}$, $\mathcal{M'}$ and $\mathcal{F}$ are
independent of the ratio of the two masses $m_1$ and $m_2$;
consequently the behavior of the relative coordinate and the stability
criterion are the same here in the test ion case ($m_1 \ll m_2$) as for
the two identical ions, which was discussed extensively in part I
\cite{GEY03}.

The eigenvalues $\lambda_{1/2}$ \eq{eq:LambdaP12} are complex for $0 <
\gamma < 1$. Then the test ion is stably synchronized with the bunch
\cite{PER82} --- which then will be stable, as the test ion describes
the motion of all the other ions in the bunch, too. This requires that
the dispersion be positive, i.e., $\alpha<1$, and the delay in the
mirror be positive, $\tau_m > 0$. For a negative delay, $\tau_m<0$,
which would lead to synchronization for $\alpha>1$, in general an
attractive interaction is necessary, which does not occur for equally
charged ions. We will in the following only consider the case of
$\alpha<1$ and $\tau_m>0$, i.e. the interaction between the test ion
and the other ions of the bunch is repulsive and it takes place around
the bunch's turning point in the mirror.

To explain the observed synchronization in the two ion model of part I
\cite{GEY03} it was sufficient that the ions experience a time delay,
when their paths cross in the mirror. The magnitude of this delay did
not matter, only its sign. However, for understanding the geometry of
the N--ion case the magnitude is important as well. In section
\ref{sec:EstimateLength} we will show that the time delay can be
obtained once one knows the bunch geometry, while the bunch geometry
in turn determines the delay. Thus, these two quantities are
determined self--consistently.

\subsection{Incorporating the effect of off--axis motion}
\label{sec:SpatialDispersion} \showlabel{sec:SpatialDispersion}

There is yet another difference between our one dimensional model and
the real trap that we need to include: in the real trap even two ions
with exactly the same energy may have different periods, as the length
of their orbits varies with the radial distance off the trap's central
axis. Pedersen \etal characterized this spread of isoenergetic periods
by a time difference $\Delta T_i$ \cite{PED01}. If this spread is too
big, synchronization is suppressed in the experiment.

\begin{figure}[t]
	\begin{center}
		\epsffile{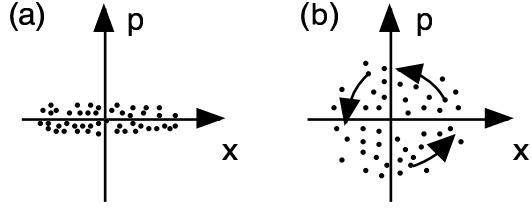}
	\end{center}
	\caption{The phase space positions of the ions in the injected
	bunch (a) serve as starting points for the ions' oscillations
	around the center of the bunch in the stable regime (b): no matter
	how small the energy (momentum) spread of the ion source is (a),
	the bunch's internal energy in the trap depends on its length.}
	\label{fig:EllipseSeed}
\end{figure}

As our one dimensional model does not allow for off--axis orbits, we
include this spread as an externally given parameter in the following
way, explained for two identical ions: In the central part $T_f$ is
effectively independent of whether a given trajectory lies on or off
the trap's central axis. So this difference in the period is
accumulated in the mirrors. There a delay $\tau_i$ of one of the two
ions with respect to the other due to the spatial dispersion leads to
a bigger separation between the ions, which is equal to extending the
time that the CM spends inside the mirror by $\frac{\tau_i}{2}$; if
the total spread after one period is $\Delta T_i=2\tau_i$ we
consequently have to replace the time interval $T_m$ in $\mathcal{M}'$
\eq{eq:MappMirrorTau} by $T_m + \frac{\Delta T_i}{4}$. This is valid
for a test ion and the bunch, too. The mirror's mapping then takes on
the following form:
\be{eq:MappMirrorMod}
	\mathcal{M}': 
	\left( \begin{array}{c} x \\ p \end{array} \right) \mapsto
	\left( \begin{array}{c} x' \\ p' \end{array} \right) = 
	\left( \begin{array}{c} x + \frac{1}{\mu}\left(T_m-\left(\tau_m -
		\frac{\Delta T_i}{4}\right)\right)p \\ p \end{array} \right)
\ee
This form reduces to equation \eq{eq:MappMirrorTau} when we define an
effective delay $\tau_m' = \tau_m - \frac{\Delta T_i}{4}$.
Consequently the stability criterion $\tau_m > 0$ for $\alpha<1$ has
to be replaced by
\be{eq:StabilityCrit}
	\tau_m' > 0 \quad \Leftrightarrow \quad 
		\tau_m > \frac{\Delta T_i}{4} \, .
\ee
The stability of the bunch is independent of the time spread $\Delta T_v$ (see
\cite{PED01}) caused by the finite energy resolution of the ion source:
when the two ions are in the synchronized mode they oscillate around
each other, exchanging momentum all the time.  Therefore even a beam
which had been injected without any energy spread is ``heated up''
corresponding to its length, as explained in reference \cite{GEY03} and
figure \ref{fig:EllipseSeed}.  The energy spread of the initial
injected bunch can be seen as a snapshot of the different ions
somewhere on their orbit around the bunch's CM.
%
\epsfxsize=6.93cm
\epsfysize=2.68cm
\showlabel{fig:EllipseSeed}

This heating is related to the time delay $\tau_m$, which we will
determine in the following section. When we then have actual numbers
for $\tau_m$ we will calculate the momentum and energy spread of the
bunch.

\section{Calculating the time delay and the bunch length}
\label{sec:EstimateLength} \showlabel{sec:EstimateLength}

From the explanations of the previous section
\ref{sec:SpatialDispersion} on the off--axis motion we conclude that
the bunching phenomenon is more stable against perturbations when the
time delay is larger: the observed bunch consequently will be the
configuration that maximizes the delay for the given experimental
parameters.

Thus, the key to understanding the properties of the N--ion bunch is
to quantify the size of the time delay $\tau_m$. This time delay
depends on the bunch geometry, while the bunch geometry and stability 
in turn depend on the time delay. Thus these two quantities have to be
determined self--consistently.

To do so we use the modified two ion model introduced in the previous
section: one of the ions with a mass $m_1=m$ is the test ion. It moves
relative to the other ion with $m_2=Nm$, describing the CM of the
other $N$ ions of the bunch. The ions' interaction, and therewith the
time delay, which we did not have to specify in the two ion case, is
now derived from the bunch geometry.

\subsection{Modified three--stage process}
\label{sec:ThreeStages}\showlabel{sec:ThreeStages}


\epsfxsize=6cm
\epsfysize=3cm
\begin{figure}[t]
	\begin{center}
		\epsffile{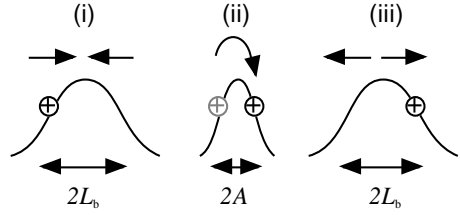}
	\end{center}
	\caption{The ions' motion in the mirror is divided into three
	stages: (i) while the bunch penetrates into the mirror the bunch
	contracts, but the ions do not change their relative ordering.
	(ii) When the bunch has reached its minimal length $A$ at the
	turning point the ions cross over the potential barrier to the
	other side and (iii) on its way out of the mirror the bunch expands
	again symmetric to (i).  The behavior of the explicitly denoted
	test ion is representative for the motion of all the other ions.}
	\label{fig:ThreeStages}
\end{figure}
\showlabel{fig:ThreeStages}

Consider first the case of non interacting ions. Then we can divide
the bunch's behavior in the mirror into three distinct steps, see
figure \ref{fig:ThreeStages}: (i) In the first stage, the bunch enters
the mirror. At this stage the ions get a compressing momentum kick,
which for each ion is proportional to its distance from the CM, see
equation \eq{eq:MomentumKick}. Given the macroscopic length of the
bunch, the momentum kick is much bigger than the spread of the ions'
momenta in the central part of the trap. Therefore, to a very good
approximation, the ions do not change their relative order, while the
bunch contracts to some minimal length, shortly before the CM turns
around in the mirror. (ii) In the second stage the CM turns around and
the ions move through the now (stationary) bunch to the other side.
(iii) In the third stage the bunch expands again on its way out of the
mirror, until the second kick at the exit of the mirror undoes the
first momentum kick and stops the expansion.

Now we have to incorporate the ions' interaction into this process.
When their interaction is weak compared to the momentum kick, this
three stage process will pertain. Since the effect of the ions'
interactions scales with their density their repulsion is most
effective when the density is the highest, i.e., when the bunch is
compressed to its minimal length at the turning point during the
second stage. We incorporate the ions' interaction into this stage
only and arrive at a modified three--stage description: the first and
last stage, where the bunch is compressed and expanded, respectively,
remain unchanged. The modification is to the second stage only, during
which the ions feel the potential of the compressed bunch and
accumulate a time delay while they move to the other side of the
bunch. Consequently the time delay and, therefore, all properties of
the stable bunch are determined by its geometry at the turning point.

\subsection{Transformation between the bunch lengths in the central part of the trap and at the turning points}

As discussed in the previous paragraph, the motion of the ions during
the second stage is determined by the momentum kick that they got at
the entrance to the mirror. The magnitude of this momentum kick is in
turn determined by the ions' respective distances from the bunch
center when the bunch enters the mirror. We therefore have to derive a
transformation for the first and third stage relating the bunch
geometry in the central part of the trap to its extension at the
turning point.

For noninteracting ions the length of the bunch in time is the same
both at the observed length $L_b$ and at the minimal length $A$. It is
identical to the time interval $\tau_k$, which defines the momentum
kick \eq{eq:tauKick}, expressed here with the average momentum $p_0$:
\be{eq:tauKickavg}
	\tau_k = \frac{L_b}{p_0 / m}
\ee
The minimal length of the bunch occurs when the CM turns around in the
mirror.  Then the first ion, being in front by $\tau_k$, has already
returned from the turning point by the distance
\be{eq:AccelByTau}
	2A = q \left( -\frac{\partial V}{\partial X}\right)_{X_m} 
		\frac{\tau_k^2}{2m} \, ,
\ee
while the last ion is still the same distance before $X_m$
\eq{eq:XmDef}. Here we assume that the mirror potential at the turning
point is linear over the (small) length $2A$ of the bunch.

In the two ion case \cite{GEY03} there was no need for some minimal
bunch length, as the crossing time of two ions is always well defined,
but for many ions the spatial extension of the continuous bunch
translates into a length in time, during which the ions cross the
bunch.

With the above form of $\tau_k$ \eq{eq:tauKickavg} 
and $E_0=\frac{p_0^2}{2m}$ we get
\be{eq:AGeneral}
	A = \frac{q}{2E_0}
		\left( -\frac{\partial V}{\partial X}\right)_{\!X_m}
		\! L_b^2 \: .
\ee
Note that the form of the trap potential between the kink and the
turning point does not influence the minimal bunch length: seen in the
laboratory frame all ions ``climb up'' the same potential ridge with
essentially the same initial energy $E_0$; therefore in our
approximation their spacing in time remains constant all the way up to
the turning point independent of the actual form of the mirror
potential. We only require a linear potential right around the turning
point.

For the following we nevertheless use the special form of our model
potential \eq{eq:trapPotential} and come back to the general form only
after we have analyzed the behavior of the observed bunch in our model
potential. Then equation \eq{eq:AGeneral} becomes:
\be{eq:ASpecific}
	A = \frac{m}{2p_0 T_m} L_b^2
\ee
With the values of the initial experiment \cite{PED01}, i.e., Ar$^+$
ions with $E_0 = 4.2$ keV and $F=80$ kV/m, a bunch of, e.g., $L_b=2$
cm is compressed to $A=0.95$ mm.

With the linear slopes at the turning point the bunch is contracted
linearly during the first stage. Therefore an arbitrary position $x$
inside the uncompressed bunch of length $L_b$ is transformed into
$x_i$ in the contracted bunch, i.e., when it has the length $A$, as:
\be{eq:PosTransform}
	x_i = \frac{A}{L_b}x \quad \Rightarrow \quad
	x_i = \frac{m L_b}{2 p_0 T_m} x
\ee

\subsection{Calculating the time delay}

According to the three stage description explained above we have to
deal with the following process: (i) the linear contraction translates
an ion from a position $x$ in the expanded bunch to the position $x_i$
in the contracted bunch \eq{eq:PosTransform}. (ii) From this position
the ion starts to cross over the now stationary bunch potential to the
other side until it reaches $-x_i$. (iii) Finally, the linear
interaction free expansion picks it up again.

Consequently the time delay $\tau_m$ due to the repulsive bunch
potential is defined as the difference between the time $T$ that the
test ion needs to cross over the bunch of minimal length $A$, starting
from an initial position $x_i$ up to the symmetric distance $-x_i$, and
$T_0$, the time needed for the same distance $2x_i$ without any
potential:
\be{eq:DelayInt}
	\tau_m = T - T_0 = \int_{x_i}^{-x_i}\!\!\!\!\!\frac{dz}{p(z)/m} 
		\; - \; \frac{2x_i}{p(x_i)/m}
\ee
Note that with the interaction the spatially symmetric positions $x_i$
and $-x_i$ are not reached at times symmetric to $T_m/2$; all ions are
delayed in the central stage, consequently they all reach their
symmetric position on the other side of the bunch later than without
interaction. The third, expanding stage only starts when all ions have
changed to their respective side of the bunch. Of course, this
separation into three successive stages is an idealization, which will
not be observed strictly in the experiment.

To calculate the times $T$ and $T_0$ we need the momentum of the ion
at the end of the first stage, i.e., at $x_i$. In addition, to
determine $T$ we need the explicit form of the bunch potential. These
quantities will be calculated in the next two subsections.

\subsubsection{Calculation of the momentum of the test ion}

To evaluate equation \eq{eq:DelayInt} we start with the momentum
$p(x_i)$: it is the sum of the ion's initial momentum relative to the
CM in the central part of the trap plus the momentum kick at the
position $x$. For a macroscopic bunch the kick is much bigger than the
momentum spread: an energy spread of 10 eV in the central part at an
ion energy of 4.2 keV corresponds to a momentum spread of about 5.7
au for Ar$^+$ ions; during the time $T_m$ this momentum difference
widens the bunch by about 0.25 mm, much less than the distance $2L_b$
of a few cm, which is traversed during the same time due to the
momentum kick. We consequently neglect the momentum spread and
calculate the ions' energy relative to the CM from the kick alone.

The momentum change of one ion relative to the bunch at the position
$x$ due to the kick was shown to be (equation \eq{eq:MomentumKick}
with $\mu=m$)
\be{} 
	\Delta p = \tau_k q F = \frac{2m}{T_m}x .
\ee
The corresponding energy relative to the bunch is $\epsilon =
\frac{\Delta p^2}{2m} = \frac{2m x^2}{T_m^2}$. With the transformation
$x = \frac{2 p_o T_m}{m L_b}x_i$ \eq{eq:PosTransform} we calculate the
energy at the contracted position $x_i$ as
\be{eq:EpsilonI}
	\epsilon (x_i) = \epsilon (x(x_i)) = 
		\frac{8 p_0^2}{m L_b^2}x_i^2
\ee
With $\epsilon(x_i)$ we can now determine $T_0$ (cf. \eq{eq:DelayInt}):
\be{eq:T0Def}
	T_0 = \frac{2x_i m}{\sqrt{2m\epsilon (x_i)}} = \frac{L_b m}{2 p_0}
\ee
$T_0$ is independent of $x_i$ because due to the linear contraction of
the bunch the momentum kick is proportional to the starting distance
$x_i$.

\subsubsection{Approximating the bunch potential}
\label{sec:BunchModelling} \showlabel{sec:BunchModelling}

To calculate the time $T$ that the test ion needs to pass over the
bunch we need the ions' distribution in the bunch. From this we
calculate the bunch potential and then the delay, which in turn
determines if this specific bunch is stable. But if it is not stable
for all $x_i$, the ions will redistribute, modifying the bunch
potential. In a more exact treatment the bunch potential would have to
be iterated so that for all ions in all parts of the bunch the
stability criterion is fulfilled.

To keep the description simple we will not allow for an arbitrary
bunch form but explicitly make the (mean field) ansatz that the ions
are spread in a Gaussian distribution around the CM all of the time,
i.e., in the central part of the trap and during the whole evolution
through the mirror. With this assumption we only need one parameter to
describe the width of the bunch. The charge density $\rho(x)$ of the
contracted bunch then will be, with the total number of ions $N$ in
the bunch and its length $A$ at the turning point:
\be{eq:GaussRho}
	\rho(x) = \frac{Nq}{\sqrt{2\pi}A}\:\rme^{-x^2/2A^2}
\ee
The ions interact via a repulsive Coulomb potential $\frac{q}{|x|}$.
To allow them to pass by each other in our one dimensional model this
singular potential is replaced by a so called ``softcore'' Coulomb
potential \cite{Softcore}:
\be{eq:SoftCoreCoul}
	W(x) = \frac{q}{\sqrt{x^2 + d^2}}
\ee
The "softcore parameter" or "impact parameter" $d$ is a measure for
how close the ions have to come when passing each other; it describes
the radius of the ion beam in the trap potential.

With this ansatz and the Gaussian density profile \eq{eq:GaussRho} we
calculate the bunch potential $W_b$ as
\be{eq:MeanBunchField}
	W_b(x_i) = \int \frac{dz\: q\: \rho(z-x_i)}{\sqrt{z^2 + d^2}}
\ee
We could now insert this potential into \eq{eq:DelayInt}, but then the
delay for a specific ion depends on where it is located in the bunch.
Therefore we concentrate for the following on the ions around the
center of the bunch. Introducing $V(x_i)$ as the quadratic expansion
of $W_b$ around the center of the bunch, we have
\be{eq:HOAppr}
	W_b(x_i) \approx V(x_i) \equiv W_b(0) - \frac{m \omega^2}{2} x_i^2
\ee
Clearly, we have
\be{eq:OmegaDef}
	 - \left. \frac{\partial^2}{\partial x^2} W_b(x)
		\right|_{x=0} = m\omega^2 \, ,
\ee
a relation which will be needed below.

With this harmonic approximation we get for the energy $E_i = E(x_i)$
of the ion above the top of the bunch potential:
\be{eq:EInitial}
	E_i = \epsilon(x_i) + V(x_i) - W_b(0) = 
	\left(\frac{8p_0^2}{m L_b^2} - \frac{m\omega^2}{2}\right) x_i^2
\ee
From this energy we may calculate the position dependent momentum as
\be{eq:MomentumOver}
	p(z) = m\omega \sqrt{\frac{2E_i}{m\omega^2} + z^2} \, .
\ee
Remember that $x_i$ denotes the starting position of the ion at the end
of the first compressing stage, just as it starts to cross the bunch
during the second stage, while the integration variable $z$ denotes
the (time dependent) position during the second, the crossing stage.

Using equations \eq{eq:DelayInt} and \eq{eq:MomentumOver} the time $T$
that the ion needs to cross over the approximated bunch potential
$V(x_i)$ \eq{eq:HOAppr} evaluates as:
\be{eq:TimeForOver}
	T = \frac{2}{\omega}\mbox{Arsinh} \left[
		\frac{L_b m \omega}{\sqrt{16 p_0^2 - L_b^2 m^2 \omega^2}}
		\right]
\ee
Here we reinserted the initial energy of the ion $E_i$
\eq{eq:EInitial} into the time dependent momentum $p(z)$
\eq{eq:MomentumOver}. Again, as for $T_0$, $T$, and therefore
$\tau_m$, is independent of $x_i$ due to the harmonic approximation
\eq{eq:HOAppr}. This approximation allows us to focus on how the delay
depends on the parameters of the trap and the bunch without having to
deal with a whole range of initial conditions $x_i$. Remember that we
already had neglected the ions' momentum against the momentum kick.

The second derivative in $m\omega^2$ \eq{eq:OmegaDef} and the
integration in $W_b$ \eq{eq:MeanBunchField} can be interchanged to
further evaluate the curvature of the bunch potential's top:
\showlabel{eq:Curvature}
\bea
	m\omega^2 
	&=& \frac{q^2 N}{\sqrt{2\pi}A^3}\int\!\!\frac{dz}{\sqrt{z^2+d^2}}
		\;\rme^{-z^2/2A^2} \left(1 - \frac{z^2}{A^2}\right) 
		\nonumber \\
	&=& \frac{q^2 N}{\sqrt{2\pi}A^3}\int\!\! dy\,\rme^{-y^2/2}
		\frac{1-y^2}{\sqrt{(d/A)^2 + y^2}}
	\label{eq:Curvature}
\eea
By changing the integration variable to $y = z/A$ the integral depends
only on the fraction $\delta = d/A$, i.e., the ratio between the beam
diameter $2d$ and the (total) length $2A$ of the bunch. In the
following we will abbreviate the integral as
\be{eq:IntDef}
	\mathcal{I}(d/A) = \mathcal{I}(\delta) =
	\int\!\! dy\,\rme^{-y^2/2}\frac{1-y^2}{\sqrt{\delta^2 + y^2}} \, .
\ee

From the expressions for $\tau_m$ according to \eq{eq:DelayInt},
\eq{eq:T0Def}, \eq{eq:TimeForOver} and \eq{eq:Curvature} we could now
calculate the time delay numerically and learn about its behavior. But
these expressions are too complex to easily ``see'' the dependencies
on the various parameters. We will therefore derive a simplified form
thereof in the following section.

\subsubsection{Linearizing $\tau_m$}
\label{sec:Linearizing}\showlabel{sec:Linearizing}

In the previous section we derived the equations necessary to
calculate the time delay numerically for a given trap configuration.
However, these equations can be simplified so that the central
stability criteria become much more obvious.

We can linearize the above set of equations by expanding both the
Arsinh and the square root in equation \eq{eq:TimeForOver}. This
approximation is valid for $L_b m\omega \ll 4 p_0$, i.e., when the
ions' energy from the momentum kick is much higher than necessary to
cross the bunch (cf. equation \eq{eq:EInitial}). Then the resulting
delay is small. From equations \eq{eq:GaussRho},
\eq{eq:MeanBunchField} and \eq{eq:OmegaDef} we see that $W_b$, and
therefore $m\omega^2$, scales linearly with $N$. When $N$ is kept
small enough this condition is fulfilled even for a short bunch or a
thin beam.

We first set $(1-x)^{-1/2} \approx 1 + x/2$ and then $\mbox{Arsinh}(x)
\approx x - x^3/6$ to approximate equation \eq{eq:TimeForOver} as:
\showlabel{eq:WurzelApprox}
\showlabel{eq:ArsinhApprox}
\bea
	T &\approx& \frac{2}{\omega}\mbox{Arsinh} \left[
		\frac{L_b m\omega}{4p_0}
			\left(1+\frac{L_b^2m^2\omega^2}{32 p_0^2}\right)
		\right] 
		\label{eq:WurzelApprox}\\
	&\approx& \frac{L_bm}{2 p_0} + \left(\frac{L_b m}{2 p_0}\right)^3
		\frac{\omega^2}{12} + \mathcal{O}(\omega^4)
		\label{eq:ArsinhApprox}
\eea
The first summand in the second line equals $T_0$ \eq{eq:T0Def}. With
equations \eq{eq:TimeInTheMirror}, \eq{eq:ASpecific},
\eq{eq:Curvature} and \eq{eq:IntDef} we finally arrive at an
expression for $\tau_m$, which depends only on parameters that
describe the bunch at the turning point in the mirror:
\showlabel{eq:TauTimesMOmega}
\showlabel{eq:TauExplizitInt}
\bea
	\tau_m &=& \frac{\sqrt{m}}{12} \left( \frac{A}{qF}\right)^{3/2}
		m\omega^2 
		\label{eq:TauTimesMOmega}\\
	&=& \frac{\sqrt{mq}}{12\sqrt{2\pi}F^{3/2}} \,\frac{N}{A^{3/2}} 
		\;\mathcal{I}(\delta)
		\label{eq:TauExplizitInt}
\eea
Finally two degrees of freedom are left, which the bunch can adjust to
achieve the most stable configuration, i.e., the biggest $\tau_m$: the
number of ions $N$ and the ratio $\delta = d/A$ between the beam width
and the bunch length at the turning point. All other quantities are
fixed in the experiment.

From \eq{eq:TauExplizitInt} it is readily seen that the delay is
proportional to the number of ions $N$ in the bunch:
\be{eq:tauLinN}
	\tau_m \propto N
\ee
The bunch can consequently stabilize itself by taking up more ions;
or, put the other way, the bunch has the tendency to keep all ions
together. It is stable with respect to the number of ions. 

\subsection{Spherical bunch geometry}

The dependence of $\tau_m$ on the beam diameter $d$ and the bunch
length $A$ at the turning point in the mirror is less obvious, but if
there is a ``most efficient'' value of $A$, i.e., one for which the
delay is maximal for a given $d$, then for this configuration the
derivative $\frac{\partial \tau_m}{\partial A}$ vanishes. Combining
all the prefactors which are not related to $d$ or $A$ into one
constant $C=\frac{N\sqrt{mq}}{12\sqrt{2\pi}F^{3/2}}$ we get from
\eq{eq:TauExplizitInt} and \eq{eq:IntDef}:
\be{eq:DTauZero}
	\frac{\partial \tau_m}{\partial A} = \frac{C}{2A^{5/2}}
		\int dy \:\rme^{-y^2/2}(y^2-1)
		\frac{3y^2+\delta^2}{\sqrt{\delta^2+y^2}^3}
\ee
It will be shown later that the extremum at $\frac{\partial
\tau_m}{\partial A} = 0$ is in fact a maximum. Since $C$ and $A$
always have a finite value, the maximum of $\tau_m$ is consequently
determined by that value of $\delta$, for which the integral vanishes.
This happens for $\delta = 1.01689\ldots$, which is about unity.
Consequently, $\tau_m$ is maximal for $A\approx d$, i.e., when the
bunch at the turning point has a spherical form. With $A=d$ the
integral \eq{eq:IntDef} in \eq{eq:TauExplizitInt} takes on a constant
value of $\mathcal{I}(1) = 0.5778\ldots$.

This value of $\delta$ of nearly unity becomes more clear when we
artificially replace for a moment the softcore interaction in the
convolution of $W_b$ \eq{eq:MeanBunchField} by a Gaussian with a width
$d$, i.e.,
\be{eq:TwoGauss}
	W_b(x_i) \simeq \frac{Nq}{2\pi Ad} \int \!\! dz \,
		\rme^{z^2/2d^2} \rme^{(z-x_i)^2/2A^2} \, .
\ee
With equation \eq{eq:OmegaDef} this gives
\be{eq:EmOmegaGauss}
	m\omega^2 = \sqrt{\frac{2}{\pi}}\; 
		\frac{Nq}{d^3 \, (1 + (A/d)^2)^{3/2}} \, .
\ee
Plugging equation \eq{eq:EmOmegaGauss} into \eq{eq:TauTimesMOmega} we 
find with the same $C$ as above that
\be{eq:TauGauss}
	\tau_m = 2C \left(\frac{A}{d^2 + A^2}\right)^{3/2} \, .
\ee
Now $\tau_m$ has its maximum where
\be{eq:DTauGauss}
	\frac{\partial \tau_m}{\partial A} = \frac{3C A^{1/2}}{(d^2 +
	A^2)^{5/2}}\;(d^2 - A^2) = 0\, .
\ee
In other words, given exactly the same functional form for both the
longitudinal bunch form and the transverse bunch potential the delay
$\tau_m$ has its maximum at exactly $A = d$. This result is not
surprising as now both ingredients of the longitudinal bunch potential
enter in a symmetric fashion, i.e., with the same functional form and
on equal footing. Then the symmetric case of their width parameters
being equal is clearly a special point; here the delay is maximal.

The slight offset to unity of the result obtained using the softcore
potential can consequently be attributed to the fact that there two
different functional forms were used for the longitudinal bunch
profile and the transverse potential. Actually, these two quantities
have different physically origins.

This result that a spherical bunch with $A=d$ leads to the biggest
delay can be understood in the following way: when the bunch is
prolate, i.e., $A>d$, the potential is flatter and slows down the ions
less, whereas in a shorter, oblate bunch the distance $2x_i$, over
which the delay is accumulated, shrinks faster than the increasing
effect of the steeper potential barrier. Of course, somewhere in
between these two extremal cases there has to be a maximum, apparently
when the bunch is spherical at the turning point.

One might argue that for very short bunches the potential barrier
should finally become so high that the ions come to a stop on top of
the barrier.  Then the delay would be high, too.  But it is not only
the length $A$ of the bunch, which determines the potential's form, but
also its width $d$: the curvature $m\omega^2$ \eq{eq:OmegaDef} at the
potential top is limited by the wider of the two functions contributing
to $W_b$ (see equation \eq{eq:MeanBunchField}, be this the Gaussian ion
distribution or the softcore Coulomb interaction.

With the more realistic softened Coulomb potential we obtained an
optimal bunch form of $A$ slightly less than $d$, but for simplicity we
will use $A=d$ in the following. As this paper focusses on the general
behavior of the bunch and not on highly accurate numbers this small
difference of a few percent is clearly negligible.

Using $A=d$, i.e., a fixed $\mathcal{I}(\delta)$, we see from equation
\eq{eq:TauExplizitInt} that the maximum delay for an otherwise fixed
trap configuration scales with the adjustable parameters $N$ and $A$
as
\be{eq:TauBehaves}	
	\tau_m \propto \frac{N}{A^{3/2}}  \propto \frac{N}{d^{3/2}}\, .
\ee
The result obtained above, that the most stable configuration is a
bunch which is spherical at the turning point, is of central
importance. As shown in part I \cite{GEY03} synchronization is an
interplay between the dispersion of the trap and the delay that the
ions experience when the bunch turns around at the turning point. The
trap dispersion is fixed by the mechanical and electrical setup of the
trap. Only the delay, which is the other central ingredient of
synchronization, can adapt. It is striking that in its most stable
configuration the bunch has a spherical form at the turning point. The
bunch observed in the central part of the trap is then the
``projection'' of this spherical bunch out of the mirrors.

To arrive at equation \eq{eq:TauExplizitInt}, which is independent of
the ions' initial conditions and is ``local in the turning point'', it
was important that both the top of the bunch potential and the ions'
energy relative to the CM had the same quadratic dependence on the
distance from the bunch's center. This is a consequence of the
constant slopes of the mirror fields. But if the mirror is not linear,
the bunch's potential at the turning point can still be expanded as in
equation \eq{eq:HOAppr}. If the bunch is not contracted linearly from
$L_b$ into $A$ in the real experiment's mirror field, then the energy
$\epsilon(x_i)$ of the ions will not be quadratic in $x_i$, but in any
case it is a symmetric function of the distance from the CM. It can
therefore at least be approximated by a quadratic dependence. Then
$\tau_m$ depends on $x_i$, but the delay is still maximal for $A
\approx d$.

\subsection{The bunch length in the central part of the trap}

With the above result, i.e., by setting $A=d$, we can use the
transformation \eq{eq:AGeneral} to predict the observed bunch length
$L_b$ as
\be{eq:LbGeneral}
	L_b = \sqrt{ \frac{8 E_0 d}{q}
		\left(-\frac{\partial V}{\partial X}\right)_{X_m}^{-1}\, .
		}
\ee
With the model potential of equation \eq{eq:trapPotential} we have
$\left(-\frac{\partial V}{\partial X}\right) = F$ and, by expressing
the ions' energy through the acceleration voltage $U$ as $E_0 = qU$,
the observed bunch length becomes
\be{eq:ObservedLb}
	L_b = \sqrt{\frac{8U}{F}d} = \sqrt{8 X_m d} \: .
\ee
For the definition of $X_m$ see \eq{eq:XmDef}. In our model trap $L_b$
is determined solely by the geometric trap parameters at the bunch's
turning point: the penetration depth $X_m$ of the ions into the mirror and
the beam width $d$ at this point. There a spherical bunch is
formed, which is then ``projected out'' into the trap's central
region. The position $X_m$ of the turning point itself is of
course determined by the ions' energy.

Note that the observed bunch length does not depend on either the mass
$m$ of the ions or on their number $N$, though the time delay
$\tau_m$ does (see equations \eq{eq:TauExplizitInt} and
\eq{eq:TauBehaves}): the reason for this unexpected behavior is that
the condition for maximal delay is that the integral in the derivative
$\frac{\partial \tau_m}{\partial A}$ \eq{eq:DTauZero} be zero --- a
condition which only depends on the geometric properties $d$ and $A$
at the turning point. More or heavier ions lead to a higher delay, but
do not change the bunch length of maximal delay.

That the turning point is of central importance for the bunching could
already be seen from the transformation \eq{eq:AGeneral}: the form of 
the potential between the field free region and the turning point has 
no effect on the relation between $A$ and $L_b$; what matters is the
slope at the turning point.

With the numerical values of the experiment, i.e., $E_0 = 4.2$ keV,
$q=+1$ and an average $F=80$ kV/m we get $X_m = 5.25$ cm. In reference
\cite{PED02b} a beam radius of $d=0.5$ mm was derived from trajectory
calculations. These trajectory calculations also had predicted a
spherical bunch at the turning point. With this value of $d$ we
predict an observed bunch length in the central part of the trap of
$L_b = 1.45$ cm. Ar$^+$ ions with the above $E_0=4.2$ keV have a
velocity of $v_0 = 14.2 \frac{cm}{\mu s}$, our $L_b$ therefore
corresponds to a length in the time domain of $W_b=0.1$ $\mu$s. For
comparison Pedersen \etal measured a value of $W_b \approx 150$ ns
(figure 9(a) of reference \cite{PED02b} for $V_1 \geq 4.5$ kV).

The greatest uncertainty in comparing our results to the measurements
stems from the greatly simplified form of our model potential. In the
experiment only the voltage on the last of the five electrodes was
varied, we can therefore not expect that the resulting field is
linear. Due to geometrical limitations, i.e., the length $L$ of the
trap is fixed and the maximal $X_m$ is limited in the experiment, it
will be difficult to check the validity of equations \eq{eq:LbGeneral}
or \eq{eq:ObservedLb} over a wide range of parameters. Also the beam
width $d$ and the slope of the mirror potential $\left(-\frac{\partial
V}{\partial X}\right)$ at the turning point have to be determined from
trajectory simulations for each field configuration.

With \eq{eq:XmDef} and \eq{eq:AlphaExp} the kinematic regime $\alpha <
1$ in which the trap has to be operated is described by the following
condition (see also equation \eq{eq:Dispersion}):
\be{eq:KinematicCond}
	E_0 > \frac{L\, qF}{4} \quad \mbox{or} \quad
	X_m = \frac{E_0}{qF} > \frac{L}{4}
\ee
The second form is a requirement on the trap's geometry: the mirrors 
have to be ``long'' enough, otherwise bunching cannot be observed.

From this condition --- which is another way of expressing the
bunching criterion $\alpha<1$ --- we see that there is a minimal bunch
length determined by the length $L$ of the field free region and the
beam diameter $2d$ at the turning point:
\be{eq:MinimalLength}
	L_b = \sqrt{\frac{2Ld}{\alpha}}
\ee
Note that though the delay increases with the number of ions in the
bunch \eq{eq:TauBehaves}, the observed $L_b$ itself is independent of
$N$ in this approximation and is determined by external parameters
only.

\section{Numerical illustration of the behavior of the time delay}
\label{sec:DelayBehave} \showlabel{sec:DelayBehave}

In Section \ref{sec:EstimateLength} we set up and determined $\tau_m$
with the quadratic approximation to the bunch potential. The result is
described by equations \eq{eq:T0Def} and \eq{eq:TimeForOver}. Then we
linearized $T$ \eq{eq:TimeForOver} and found our central result that
the time delay is maximal for $A=d$. Now we go back to equations
\eq{eq:T0Def} and \eq{eq:TimeForOver} and numerically calculate the
time delay from them. This will both confirm our findings with the
linearized $T$ and illustrate the behavior of the bunch with respect
to the most important parameters.

The delay depends on a number of parameters, some of which are
``externally given'', i.e., fixed for the given experimental setup,
while others describe how the bunch ``reacts'' to this environment.
For instance, the size $L$ of the trap, the energy $E_0$, the charge
$q$ and the mass $m$ of the ions and the slope of the mirror fields
$F$ are fixed. Derived from those are the momentum $p_0 =
\sqrt{2mE_0}$, the time $T_m$ \eq{eq:TimeInTheMirror} and the
dispersion parameter $\alpha$ \eq{eq:AlphaExp}, which we consider
fixed, too. In the following these quantities will be set to their
respective experimental values of $E_0 = 4.2$ keV, $F=80$ kV/m, $q=+1$
e and $m=40$ amu, which were used in the initial experimental
discovery of the bunching effect \cite{PED01}. With these values we
get $T_m = 1.48$ $\mu$s and $\alpha = 0.956$.

On the other hand we have the number of ions in the bunch $N$, the
observed length $L_b$ and the length $A$ in the mirror. There is no
constraint on these quantities, so the bunch will try to adjust them
to achieve the most stable configuration possible in the given trap
regime. The two lengths, $L_b$ and $A$, are not independent, though.
They are connected to each other by the transformation
\eq{eq:AGeneral}, determined by the slope of the mirror potential at
the turning point. We view the minimal length in the mirror, $A$, as
the more fundamental of these quantities, characterizing the minimal
width in the mirror region where the ions cross the bunch; the
experimentally observable $L_b$ is merely the projection of this
contracted length out of the mirror. We do not expect the absolute
values which we predict for the observed $L_b$ to be exact, but we
expect that the important dependencies, governed by $A$, will be
reproduced.

The parameters we mentioned up to now are easily classified as fixed
or variable. The beam radius $d$ has a special role: it belongs to the
fixed quantities, as it is determined by the electrostatic potentials
on the mirror electrodes, but in our treatment it appears as a
constituent of the bunch potential $W_b$ \eq{eq:MeanBunchField} and it
is closely intertwined with $A$, as we have seen above. We already
know from equation \eq{eq:DTauZero} that the delay is maximal for $A
\approx d$. Nevertheless we will treat $d$ as a variable parameter in
the following numerical illustrations --- just to see our results of
the previous section confirmed.

We will now in turn vary each of these three parameters, $A$, $d$ and
$N$, while the other two remain fixed. This will give three different
views on the behavior of the bunch.

\subsection{Time delay vs. bunch length $A$ and validity of the approximations}
\label{sec:TauVSA} \showlabel{sec:TauVSA}

\epsfxsize=8cm
\epsfysize=6cm
\begin{figure}[t]
	\begin{center}
		\epsffile{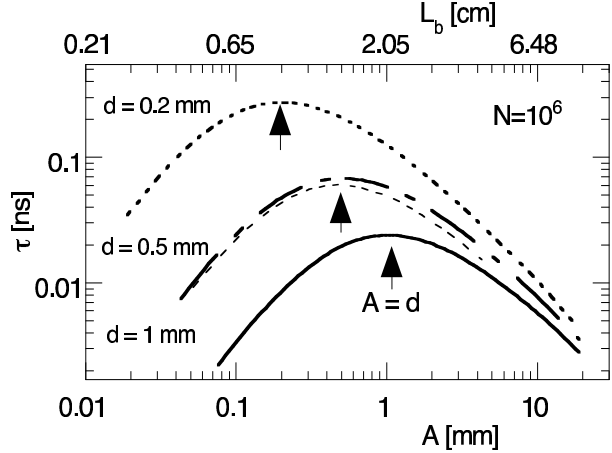}
	\end{center}
	\caption{Time delay calculated from\eq{eq:T0Def},
	\eq{eq:TimeForOver} and \eq{eq:Curvature} for a constant number of
	ions $N=10^6$ and fixed beam diameters of $d=$ 0.2 mm (- - -), 0.5
	mm (--- - ---) and 1 mm (------). The arrows indicate for each
	curve where the bunch length is equal to the beam diameter, i.e.,
	$A=d$. The thin broken curve plots for $d=$ 0.5 mm the weighted
	average of the delay from the equations of motion with the full
	bunch potential, i.e., without the harmonic approximation
	\eq{eq:HOAppr} (see text).}
	\label{fig:tauA}
\end{figure}
\showlabel{fig:tauA}

First we look at the time delay as a function of the bunch length $A$
for fixed $N$ and $d$. We evaluate equations \eq{eq:T0Def},
\eq{eq:TimeForOver} and \eq{eq:Curvature} with $N=10^6$ for three
values of $d$ --- 0.2 mm, 0.5 mm and 1 mm --- and plot the results in
figure \ref{fig:tauA}. For short bunches, i.e., small $A$, and for
long bunches the delay is small with a maximum in between. This
maximum of $\tau_m$ occurs around $A=d$, where the length and the
width of the bunch are equal. This is exactly the behavior predicted
in the previous section.

The linearized result \eq{eq:TauExplizitInt}, from which we derived
the condition $A=d$, is indistinguishable from the full solution in
figure \ref{fig:tauA}; it differs by less than one percent for small
$A$ and coincides for larger $A$. This highlights the fact that with
these parameters the delay of the ions through the bunch is a small
perturbation compared to the momentum kick from the mirror potential.

\epsfxsize=8cm
\epsfysize=5cm
\begin{figure}[t]
	\begin{center}
		\epsffile{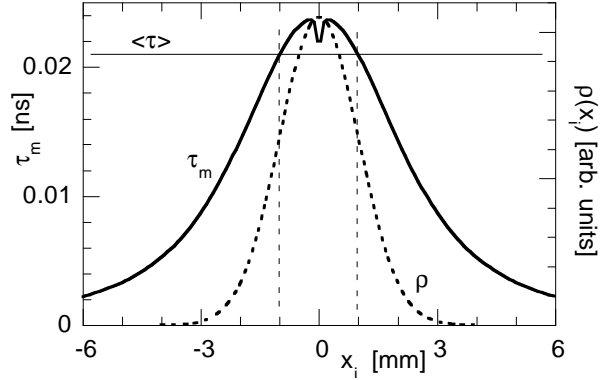}
	\end{center}
	\caption{Time delay $\tau_m(x_i)$ computed from the equations of
	motion (solid line) and the ion density $\rho(x_i)$ of the bunch
	at the turning point (broken curve) against the initial distance
	$x_i$ \eq{eq:PosTransform}, both for $A=d=1$ mm (cf. equation
	\eq{eq:GaussRho}). The solid horizontal line denotes the weighted
	average $\langle \tau \rangle$ of $\tau_m$, while the two vertical
	broken lines at $x_i = \pm A$ mark the width of the ion bunch.}
	\label{fig:tauVert}
\end{figure}
\showlabel{fig:tauVert}

To check the validity of our harmonic approximation of the bunch
potential \eq{eq:HOAppr} we numerically evaluated $\tau_m$ from
equation \eq{eq:DelayInt}. This means that we still keep the modified
three--stage model of section \ref{sec:ThreeStages}, where the
interaction between the test ion and the bunch is confined to the
second stage, during which the bunch is static. The position dependent
momentum $p(z)$ in equation \eq{eq:DelayInt} is calculated from its
energy from the momentum kick $\epsilon(x_i)$ \eq{eq:EpsilonI} and the
bunch potential $W_b(x_i)$ of equation \eq{eq:MeanBunchField} as
\be{eq:MomentumZ}
	p(z) = \sqrt{2 E(z) m}\quad \mbox{with} \quad
		E(z) = \epsilon (x_i) - (W_b(0) - W_b(z))\, .
\ee
$E(z)$ is the energy of the ion above the bunch potential. The
resulting delay $\tau_m = \tau_m(x_i)$ now depends on the ion's
position in the bunch $x_i$ \eq{eq:PosTransform} at the end of the
first stage, during which the bunch was compressed.

In \fig{fig:tauVert} the delay $\tau_m$ is plotted against the
starting position $x_i$ for $A = d = 1$ mm. For comparison also the
contracted longitudinal bunch profile $\rho(x_i)$ is shown. We see
that $\tau_m$ decays much slower with increasing distance $x_i$ from
the bunch's center than $\rho(x_i)$: at $x_i = 2.5$ mm, where the
bunch density becomes negligible, $\tau_m$ is still half of the
maximal value close to the center. Thus most of the ions have
a $\tau_m$ of about the same magnitude. Consequently, the density
weighted average
\be{eq:AvgTau}
	\langle \tau \rangle = \int \!\! dx_i\; \rho(x_i)\, \tau_m(x_i)
\ee
is a reasonable approximation over the whole length of the bunch. Also
for different values of $d$ the ratio between the widths of $\rho$ and
$\tau$ remains unchanged, figure \ref{fig:tauVert} is a representative
example.

The density averaged delay $\langle \tau \rangle$ for $d$ = 0.5 mm is
given in figure \ref{fig:tauA} as a thin broken curve. It is quite
close to the result with the harmonic approximation to the bunch
potential. Thus, both the approximation to $W_b$ \eq{eq:HOAppr} and
the subsequent linearization of $T$ (\eq{eq:WurzelApprox} and
\eq{eq:ArsinhApprox}) are justified. For the following illustrations
we will therefore continue to evaluate equations \eq{eq:T0Def},
\eq{eq:TimeForOver} and \eq{eq:Curvature}.

\subsection{Time delay vs. beam diameter $d$}
\label{sec:TauVSD} \showlabel{sec:TauVSD}

\epsfxsize=8cm
\epsfysize=5cm
\begin{figure}[t]
	\begin{center}
		\epsffile{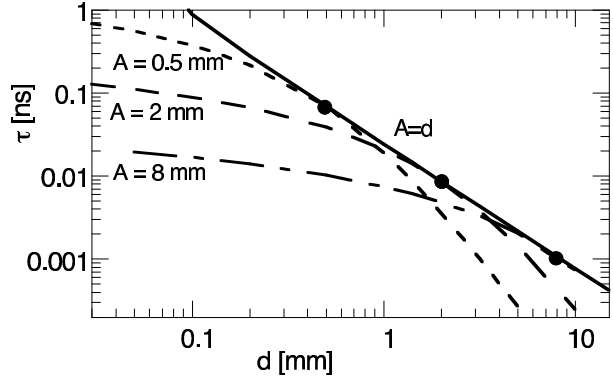}
	\end{center}
	\caption{Time delay as a function of the beam diameter for
	$N=10^6$ and various fixed bunch lengths of $A = 0.5$ mm (- - -),
	2 mm (--- ---) and 8 mm (--- - ---). The enveloping solid line
	marks the time delay for $A=d$ according to \eq{eq:TauBehaves};
	the dots indicate the diameters, which correspond to the three
	bunch lengths shown. These correspond to values of $L_b$ of 14, 29
	and 58 mm, respectively.}
	\label{fig:tauD}
\end{figure}
\showlabel{fig:tauD}

When we now keep the bunch length fixed and vary the beam diameter the
maximum of the delay for a spherical bunch shows up again. In
\fig{fig:tauD} the delay is plotted for three different bunch lengths.
It decreases with increasing beam diameter, but for fixed $d$ the
maximum is at $A=d$. This configuration is indicated by the solid
enveloping line with a slope of $-3/2$ according to equation
\eq{eq:TauBehaves}.

From this plot it can be seen, too, that when the spatial dispersion
requires a minimal delay of, e.g., $\Delta T_i/4$ = 0.025 ns (see
\eq{eq:StabilityCrit}) then the beam has to be focussed to a radius of
no more than 1 mm.

The necessary radial focussing can also be determined by inserting
$A=d$ and $4 \tau_m > \Delta T_i$ \eq{eq:StabilityCrit} into equation
\eq{eq:TauExplizitInt}. The beam radius in millimeters has to be
smaller than
\be{eq:minimalD}
	d [\mbox{mm}] < 5.02\times 10^{-4} 
		\frac{(mq)^{1/3}N^{2/3}}{F\,\Delta T_i^{2/3}}\, ,
\ee
when the mass $m$ is given in atomic mass units (amu), the charge $q$
in elementary charges, the slope $F$ in kV/m and the dispersion
$\Delta T_i$ in nanoseconds.

\subsection{Time delay vs. ion number $N$}
\label{sec:maxDelay}\showlabel{sec:maxDelay}

\epsfxsize=8cm
\epsfysize=8.7cm
\begin{figure}[t]
	\begin{center}
		\epsffile{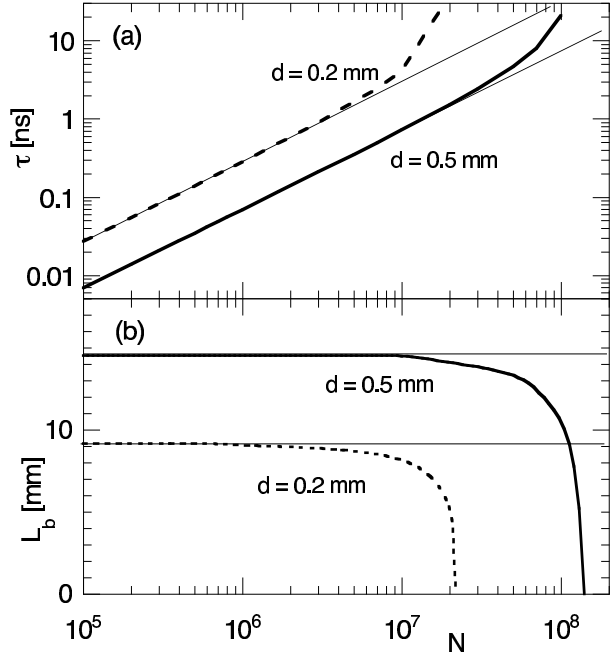}
	\end{center}
	\caption{(a) Maximal time delay achievable at a given beam
	diameter and (b) the observed bunch length $L_b$, for which this
	maximal $\tau_m$ is reached, as a function of the number of ions
	$N$. The thin solid lines indicate in (a) the linear relation
	between $\tau_m$ and $N$ \eq{eq:tauLinN} and in (b) the constant
	$A=d$ according to the linearized solution of section
	\ref{sec:Linearizing}.}
	\label{fig:tauA-N}
\end{figure}
\showlabel{fig:tauA-N}

Another view of the behavior of the time delay is presented in figure
\ref{fig:tauA-N}. Now for given $N$ and $d$ the bunch length $A$ is
optimized to maximize the time delay. Panel (a) plots the maximal
delay according to the optimized $A$. Panel (b) gives the
corresponding observed bunch length $L_b$. Both the maximal $\tau_m$
and $L_b$ are plotted vs. $N$ for two different values of $d$.

For small $N$ the delay grows linearly with $N$, as already derived in
equation \eq{eq:TauBehaves}. This behavior is depicted by the thin
solid lines of slope 1 in panel (a). In this linear regime the bunch
is spherical, as can be seen in panel (b): $L_b$ is constant, which
implies that $A$ is constant, too (see \eq{eq:ASpecific}). The thin
solid lines in panel (b) mark the values of $L_b$, that result from
inserting the two beam radii $d$ = 0.5 mm and $d$ = 0.2 mm into
\eq{eq:ASpecific}.

For higher ion numbers, however, the maximal achievable delay
increases (panel (a)) and is reached for shorter and shorter bunches
(panel (b)). Finally, the ion number, and therefore the density of the
bunch, will become so high that the ions cannot pass over the bunch
any more with the energy from the momentum kick. The ions then bounce
back from the bunch and do not reach the other side any more. This is
where $4 p_0 \leq L_b m \omega$, consequently our formulation for the
crossing time $T$ \eq{eq:TimeForOver} breaks down.

Inserting equations \eq{eq:Curvature}, \eq{eq:ASpecific} and
\eq{eq:TimeInTheMirror} into this condition $4p_0 \geq L_b m \omega$
for \eq{eq:TimeForOver} to be valid we see that this determines the
maximal number $N_{\maxi}$ as
\be{eq:MaxAnzahl}
	N_{\maxi} = \frac{4\sqrt{2\pi}}{\mathcal{I}(1)}\, \frac{d^2 F}{q}\, .
\ee
Note that $N_{\maxi}$ is proportional to the beam area via $d^2$:
this is consistent with our previous explanation that there is a
maximal potential barrier from the contracted bunch that the ions can
overcome with the energy from the momentum kick.

Consequently there are --- for a given $d$ --- two limits for the
number of ions in a bunch: the lower limit is determined by the minimal
delay necessary to compensate for the spatial dispersion $\Delta T_i$
and the maximal $N$ stems from the requirement that the ions have to be
able to pass over the contracted bunch's potential.  

Equation \eq{eq:minimalD} in the previous section can be solved for
the minimal number of ions necessary to support a given spatial
dispersion $\Delta T_i$:
\be{eq:MinAnzahl}
	N_{\mini} = \frac{3\sqrt{2\pi}}{\mathcal{I}(1)}\;
		\frac{d^{3/2} F^{3/2}}{\sqrt{mq}}\; \Delta T_i
\ee
With the numbers from the experiment we calculate from
\eq{eq:MaxAnzahl} and \eq{eq:MinAnzahl} that for a beam radius of
$d=0.5$ mm and a dispersion of $\Delta T_i = 0.1$ ns \cite{PED01} the
number of ions may vary in the range of $N=3\times 10^5 \ldots 2\times
10^8$, i.e., over nearly three orders of magnitude. Within this range
the bunch length is independent of the number of ions.

This is consistent with the experimental observation that the bunch
length is essentially independent of the density of the injected ion
beam over at least three orders of magnitude \cite{PED01}. Our
estimate is more restrictive, especially $N_{\maxi}$ is bigger in the
experiment. The reason lies in our harmonic approximation for
$\tau_m$: as we see from figure \ref{fig:tauVert} the delay is smaller
for ions in the wings of the bunch, which means that for them the
momentum kick increases faster than the height of the potential
barrier. These ions will be able to pass over a bunch, which already
reflects the ions close to its center. These will move away from the
center of the bunch. Consequently the form of the bunch changes,
allowing the bunch to be stable for higher ion densities at the
turning point than we estimated here. In order to find the ``true''
$N_{\maxi}$ a dynamical model with a truly self consistent bunch
potential has to be used; our fixed Gaussian form of the density is
too simplified to give accurate numbers in this extreme regime.

\section{Momentum spread of the bunch}
\label{sec:PhaseSpace}\showlabel{sec:PhaseSpace}

In section \ref{sec:SpatialDispersion} we stated that the bunch is
heated up to a specific energy spread in the trap by the momentum kicks
from the trap potential, no matter how small the initial energy spread
at injection time had been.

In ``part I'' we derived a relation between the maximal distance between
the two ions $x_{\maxi}$ and their maximal relative momentum
$p_{\maxi}$ in the central part of the trap. As explained in section
\ref{sec:SpatialDispersion} in the real trap the effective delay
$\tau_m' = \tau_m - \Delta T_i/4$ has to be used, and therefore
equation (37) of reference \cite{GEY03} becomes
\be{eq:Beta}
	p_{\maxi} = \frac{x_{\maxi}}{\beta} 
	\quad \mbox{with} \quad
	\beta = \frac{T_m}{2 m}\sqrt{\frac{T_m(1-\alpha)}{\tau_m'} - 
	(1-\alpha)^2}\, .
\ee
This relation is valid for a constant $\tau_m'$. Our harmonic
approximation to the bunch potential leads to the same constant delay
for all ions of the bunch; consequently we can use this relation to
relate the longitudinal momentum spread of the bunch $\Delta p$ to its
extension in the central part of the trap $L_b$ as $\Delta p = L_b /
\beta$.

With the parameters of the original experiment, i.e., $T_m = 1.48$
$\mu$s, $\alpha = 0.956$ and $\tau_m' \leq 0.1$ ns, the second term in
the square root in the above equation \eq{eq:Beta} is smaller than the
first by six orders of magnitude and can therefore be neglected. We
see then, that $\Delta p$ scales with the square root of the effective
time delay $\tau_m'$ and is proportional to the bunch length $L_b$:
\be{eq:DEvonTau}
	\Delta p = \frac{2 m}{T_m}
		\sqrt{\frac{\tau_m'}{T_m(1-\alpha)}} \; L_b
\ee
This form highlights that there are two different routes from the
stable regime to the stability limit. One route is to increase the
slope of the mirror potential to change the dispersion of the trap so
that $\alpha \to 1$. In this case the momentum spread grows only
slowly as long as the trap is ``far'' in the stable regime, but as
the mirror slope $F$ approaches the maximal value $F_{\maxi} = \frac{4
E_0}{L q}$ (equation \eq{eq:AlphaExp} with $\alpha=1$) $\Delta p$
increases sharply and the bunch becomes drastically ``hotter''. Right at the
stability limit $\alpha=1$ we get $\Delta p = \infty$. The bunch is 
``blown apart'' by its internal energy. Correspondingly
the notion of a ``momentum spread of the bunch'' loses its meaning,
because the bunch itself ceases to exist. This ``heating'' of the 
bunch is a collective effect of all ions together.

The other route, which was discussed less extensively, is to relax the
radial focussing of the beam to allow for stronger off--axis motion of
the ions. This increases $\Delta T_i$, and therefore decreases
$\tau_m'$. Then $\Delta p$ decreases with $\Delta T_i/4 \to \tau_m$,
i.e., the bunch becomes ``cooler'' closer to the stability limit. This
happens because a stronger off--axis motion increases the spread of
the periods of the individual ions through the trap, which effectively
stretches the bunch in time and consequently decreases the interaction
between the ions at the turning point. In contrast to the first
collective route to instability via the \emph{longitudinal} trap
dispersion this second effect stems from the \emph{transverse}
dispersion due to different off--axis orbits of the individual ions.
On the first route the bunch is ``blown apart'' because the bunching
mechanism cannot confine the internal energy any more, while on the
second route the bunch ``boils off'' ions that stray away too far from
the bunch. Only those ions that have similar orbits through the trap
stay together, thus reducing the internal momentum spread of the
remaining bunch.

We once again can insert the numbers from the original experiment, i.e.,
$T_m = 1.48$ $\mu$s, $F = 80$ kV/m, $q=+1$ and $\alpha = 0.956$ and
find from figure \ref{fig:tauA} that $\tau_m = 0.07$ ns at $d=0.5$ mm
for $N=10^6$ ions. With a realistic $\Delta T_i = 0.18$ ns
\cite{PED01} and $x_{\maxi} = L_b = 1.44$ cm we calculate a momentum
spread of $\Delta p = 13.5$ au. With $m=40$ amu this translates into
an energy spread in the laboratory frame of $\Delta E = \frac{p_0}{m}
\Delta p = 0.87$ au = 24 eV at a beam energy of $E_0 = 4.2$ keV.

\section{Comparing to the empirical criteria for bunching}
\label{sec:Criteria} \showlabel{sec:Criteria}

In reference \cite{PED02b} three empirical criteria were given by
Pedersen \etal for synchronization to occur. We now show how these
criteria are related to the results derived from our mapping approach
about how and when synchronization occurs and about the size and
behavior of the bunch.

(i) The first criterion is the so called "kinematical criterion" which
in \cite{PED02b} is given as $\frac{\partial T}{\partial E} > 0$. This
criterion is obviously the same as our stability criterion of $\alpha
< 1$ for $\tau_m > 0$, see equations \eq{eq:Dispersion} and
\eq{eq:LambdaP12}. Both expressions state that the dispersion of the
trap has to be positive, i.e., that a faster ion has a longer period.

(ii) The second criterion of Pedersen \etal, the "focussing
criterion", states that the interaction of the ions is important only
in the mirror regions and that the beam diameter has to be compressed
to about the same width as its length. We have shown in part I
\cite{GEY03} that collisions in the central part are far more unlikely
than collisions in the mirror and that only collisions in the mirror
can synchronize the ions. We found the requirement of comparable
longitudinal and transverse dimensions at the turning point when we
estimated the bunch length in sections \ref{sec:EstimateLength} and
\ref{sec:DelayBehave}: a spherical bunch is most efficient in delaying
the ions and therefore most stable against external perturbations.
With this result, i.e., $A=d$, and the transformation \eq{eq:AGeneral}
between $A$ and $L_b$ we then derived an estimate for the observed
bunch length \eq{eq:LbGeneral}, which agrees with the prediction of
Pedersen \etal (see equation (29) of reference \cite{PED02b}).

(iii) The third criterion of Pedersen \etal, the "collision
criterion", states ``[\ldots] the collision probability at the turning
point must ensure that the ions indeed lock their motion [\ldots]''
and ``[\ldots] too few collisions and too many lead to diffusion
[\ldots]''. In our treatment the ``number of collisions'' corresponds
to the magnitude of the effective time delay $\tau'_m = \tau_m -
\frac{\Delta T_i}{4} > 0$ (see equation \eq{eq:StabilityCrit}). As
discussed in sections \ref{sec:TauVSD} and \ref{sec:maxDelay} the
delay $\tau_m$ due to the collisions has to be bigger than the
intrinsic dispersion $\Delta T_i/4$ for synchronization to survive
this perturbation, corresponding to the observation that too few
collisions lead to diffusion. We want to emphasize again that the time
spread $\Delta T_v$ of reference \cite{PED02b} due to the ions'
different energies is not a perturbation but a part of the ion
dynamics.

The upper limit of too many collisions, was illustrated in
\fig{fig:tauA-N}: if the bunch contains too many ions or is focussed
too tightly, the ions bounce off the bunch potential, experience a
negative delay and diffuse out of the bunch. In the framework of our
approximations this upper limit is described by equation
\eq{eq:MaxAnzahl}.

\section{Summary and conclusions}
\label{sec:summary} \showlabel{sec:summary}

In this paper we extended a recently proposed description of the
synchronization effect in the ``Zajfman trap'' \cite{GEY03} from
describing the motion of two identical ions to the behavior of a
macroscopic ion bunch: the motion of the two identical ions through
the trap field was described in their CM system by three mappings, one
for each of the different parts of the trap. When the paths of the
ions cross in the mirror their repulsive interaction slows them down
temporarily and the resulting time delay, together with the dispersion
of the trap, is responsible for coupling their motion. Here we
generalize this two ion approach to describe the motion of a
(representative) test ion in a macroscopic bunch. The same stability
condition applies as in the two ion case. The interaction between the
ions is now specified through the size and form of the bunch. Together
with a relation between the observed bunch length in the central part
of the trap and its size at the turning point in the mirror the time
delay can be calculated explicitly.

We show that the time delay is maximal for a bunch which is spherical
at the turning point and we calculate how the observed bunch length
derived from this most stable configuration depends on the various
parameters of the system.

Not only does our prediction of the observed bunch length match the
experimental results within the level of approximation of our model,
but we also confirm that the bunch length is essentially independent
of the number of ions. The lower and the upper limits to this number
are identified: the lower limit is connected to the spread of the
ions' periods due to off--axis motion, as the time delay that the ions
experience when they cross the bunch has to be big enough to
compensate for it. At the upper limit the repulsive bunch potential of
the compressed bunch at the turning point becomes too high and the
ions cannot cross the bunch any more. Then they bounce back from the
bunch, the resulting time delay changes its sign, the ions do not
synchronize any more and consequently the bunch loses ions.

The numerical evaluation of the time delay for various parameters
illustrates the behavior of the bunch and confirms the validity of the
approximations employed in of our description. We make several new
predictions that we hope will be tested experimentally: 
\begin{itemize}
	\item[(i)] To directly observe our central result of $A=d$ at the
	turning point some kind of imaging technique has to be used. This
	would allow one to capture the form of the bunch at that moment
	when it comes to rest without having to resort to trajectory
	calculations.

	\item[(ii)] Much easier than the direct visualization of the
	turning bunch is to check the validity of equation
	\eq{eq:ObservedLb}, which is based on $A=d$. For a constant mirror
	slope $F$ at the turning point we predict that the bunch length
	observed in the central part of the trap $L_b$ scales with the
	square root of the beam diameter $2d$ at the turning point. For a
	fixed $d$, $L_b$ would be inversely proportional to the square root
	of the slope $F$. However, $L_b$ cannot become shorter than a
	minimal value, which depends on the total length $L$ of the field
	free region of the trap according to \eq{eq:MinimalLength}.

	\item[(iii)] We found that the bunch length is determined by the
	mirror slope at the turning point only and not by the dispersion,
	see \eq{eq:LbGeneral}. This can be verified with a nonlinear
	mirror potential, which keeps the slope at the turning point
	constant for different trap dispersions. 
	
	\item[(iv)] The observed bunch length is independent of the number
	of ions in the bunch within certain limits, which are given by
	equations \eq{eq:MaxAnzahl} and \eq{eq:MinAnzahl}. The two limits
	scale differently with the parameters of the trap and the ions.
	
	\item[(v)] With equation \eq{eq:DEvonTau} we derived a relation
	between the bunch size and its longitudinal momentum spread.
	According to this equation there are two ways that a bunch can
	become unstable, either via the longitudinal or via the transverse
	dispersion of the trap. The two routes, depending on the mirror
	slope and the beam diameter, respectively, have opposite effects
	on the momentum spread.
\end{itemize}

Finally we explain the empirical conditions for synchronization given
by Pedersen \etal in the framework of our model.

In this paper we considered only a static bunch, i.e., the bunch
exists initially and is either stable or not. Future work therefore
has to focus on the dynamical properties of the synchronization
effect, as, e.g., how and at which rate the bunch acquires additional
ions from which parts of phase space. Other open questions are related
to the issues of ions of different masses: what is the mass resolution
of a ``Zajfman trap'' operated as a mass spectrometer? Do bunches of
different ions ``stick together'' or can one have more than one bunch
in the trap independently?

Another set of questions is connected to the internal dynamics of the
bunch and whether there is an application for this ``inner trap''
moving through the outer macroscopic trap.

It is not clear at the moment if our simple model can be used to
answer these questions or if a more detailed description of the
many--ion dynamics has to be used. Nevertheless, our simple model can
explain the underlying mechanism and the stability criteria and
provide insight into how the macroscopic bunch reacts to the
externally given conditions and parameters.

\section*{Acknowledgments}

We thank Daniel Strasser and Daniel Zajfman for constructive
discussions and further explanations of the experiment.

This research was funded by the Israel Science Foundation.



\section*{References}

\begin{thebibliography}{99}
	\bibitem{PED01} H~B~Pedersen \etal,
	\PRL \textbf{87} (2001) 055001

	\bibitem{ZAJ97} D~Zajfman \etal,
	\emph{Phys. Rev. A} \textbf{55} (1997) R1577
	
	\bibitem{STR02} D~Strasser \etal,
	\PRL \textbf{89} (2002) 283204
	
	\bibitem{MAR98} A~G~Marshall, C~L~Hendrickson and G~S~Jackson,
	\emph{Mass. Spec. Rev.} \textbf{17} (1998) 1
	
	\bibitem{ATT05x} D~Attia \etal,
	arXiv:physics/0503117
	
	\bibitem{LAW88} J~Lawson,
	\emph{The Physics of Charged Particle Beams}, 
	Clarendon Press, Oxford, 1988, 2nd edition
	
	\bibitem{GEY03} T~Geyer and D~J~Tannor,
	\jpb \textbf{37} (2004) 73
	
	\bibitem{PED02b} H~B~Pedersen \etal,
	\emph{Phys.~Rev.~A} \textbf{65} (2002) 042704
	
	\bibitem{PED02a} H~B~Pedersen \etal,
	\emph{Phys.~Rev.~A} \textbf{65} (2002) 042703
	
	\bibitem{PER82} see, e.g., I~Percival and D~Richards, 
	\emph{Introduction to Dynamics},
	Cambridge University Press, Cambridge, 1982
	
	\bibitem{Softcore} R~Grobe, J~H~Eberly,
	\emph{Phys.~Rev.~A} \textbf{48} (1993) 4664

	
\end{thebibliography}
\end{document}